\documentclass[prd,preprint,superscriptaddress,nofootinbib,longbibliography,aps]{revtex4-1}
\usepackage[utf8]{inputenc}
\usepackage{graphicx}
\usepackage{longtable}
\usepackage{amsmath}
\usepackage{color}
\usepackage{amssymb}
\usepackage{subfigure}
\usepackage{tabularx}
\usepackage{microtype}
 \usepackage[linktoc=all]{hyperref}
 \usepackage{cleveref}
\usepackage{stackengine}
\usepackage[normalem]{ulem}
\usepackage{slashed}
\usepackage{color}
\usepackage{bm}
\usepackage{braket}
\usepackage{slashed}
\usepackage{comment}
\definecolor{myblue}{rgb}{ 0.188, 0.478,0.858}

\newcommand{\odp}{ \hat \Omega\cdot \hat p}

\begin{document}  
\title{Testing Gravity with Frequency-Dependent Overlap Reduction Function in Pulsar Timing Array}

\author{Qiuyue Liang}
\email{qiuyue.liang@ipmu.jp}
\author{Ippei Obata} 
\email{ippei.obata@ipmu.jp}
\author{Misao Sasaki}
\email{misao.sasaki@ipmu.jp}

\affiliation{Kavli Institute for the Physics and Mathematics of the Universe (WPI), University of Tokyo,
Kashiwa, Chiba 277-8583, Japan} 

\date{\today}

\begin{abstract} 
The positive evidence of a nano-hertz gravitational wave background recently found by several pulsar timing array (PTA) collaborations opened up a window to test modified gravity theories in a unique frequency band in parallel to other gravitational wave detection experiments. 
In particular, the overlap reduction function (ORF) in PTA observation is sensitive to the phase velocity of gravitational waves. 
In this work, we provide analytical expressions for the coefficients of the multipole moments in the ORF, and utilize these analytical results to study constraints on
the phase velocity from the frequency dependent overlap reduction function obtained from the Chinese PTA (CPTA) data. 
While the data contain large error bars yet, interesting constraints are found in the frequency-dependent ORF in the case of subluminal phase velocity. This makes us expect that the nano-hertz band gravitational wave background will become one of the important arenas for exploring modified gravity theories. 
\end{abstract} 
\preprint{YITP-24-55}
\maketitle

\section{Introduction} 
Pulsars are compact neutron stars emitting pulsed lights with very stable rotation periods \cite{Hewish:1968bj}.
Thanks to this feature, their pulse arrival time measurements can test the fluctuations in the gravitational field at frequencies on the order of 1-100nHz \cite{Detweiler:1979wn} with long temporal baseline data. 
Moreover, we can construct the correlation function among the pulsar pairs in the system to study the underlying gravitational wave background (GWB). 
For an isotropic GWB, we can split the signal into power spectrum and the overlap reduction function, which is a characteristic angular correlation function between pulsar pairs, dubbed the Hellings-Downs (HD) curve  \cite{Hellings:1983fr}.
Recently, an evidence of a nano-hertz GWB in the pulsar timing array signals, over 3 $\sigma$ statistical significance, has been reported by NANOGrav, PPTA, EPTA, and CPTA \cite{NANOGrav:2023gor,NANOGrav:2023hvm,EPTA:2023fyk,Reardon:2023gzh,Xu:2023wog}.
This unique probe opened up the possibility to test fundamental theories, including modifications to general relativity.  

Constraining modified gravity theories with extra non-Einstein polarization modes and/or modified dispersion relations by using PTA has been well discussed in the literature \cite{Lee:2010cg,Gair:2015hra,Qin:2020hfy,Liang:2021bct,Liang:2023ary,Bernardo:2022rif,Bi:2023ewq,Bernardo:2023zna,Wu:2023rib,Wu:2023pbt}. 
Constraining the graviton mass is, among others, of particular research interest \cite{Finn:2001qi,LIGOScientific:2016lio,Liang:2021bct,LIGOScientific:2021sio,Bernardo:2023mxc,Wang:2023div,Wu:2023rib,Bernardo:2023zna}. 
A modified dispersion relation of gravitational waves will give rise to a deviation in  the overlap reduction function from the HD curve. 
As pointed out in \cite{Liang:2023ary}, the phase velocity other than the group velocity $v_p=\omega(k)/k$ is the parameter entering the observable, where $\omega(k)$ and $k$ are the angular frequency (or energy) and the wavenumber (or momentum) of the gravitational wave, respectively. While we have $v_p>1$ in canonical massive gravity, theories with $v_p <1$, which we refer to as subluminal gravity, can naturally arise from various effective field theories of gravity \cite{CarrilloGonzalez:2022fwg,Ezquiaga:2021ler,deRham:2019ctd}.
In this paper, we derive analytical expressions for the coefficients of the multipole moments of ORF for both $v_p>1$ and $v_p <1$ cases. 

It is of great interest to apply these analytical expressions to the current data sets and see if we can obtain constraints on the phase velocity. 
In this respect, as one generically expects that the dispersion relation is frequency dependent, CPTA offers a unique opportunity in that they report an analysis of the overlap reduction function based on three different frequency bins, in contrast to those by the other experimental groups that provide only frequency-averaged data analyses.
In this paper, as an application of our analytical formula, we discuss the frequency-dependent angular correlation obtained from CPTA.
We mention, however, CPTA has been operating only for 3.4 years, and only one frequency bin, $f= 1.5/(3.4\,{\rm yrs})$, shows significant statistical evidence of a HD signal. Therefore, we will focus on this specific frequency, while hoping for the possibility that future PTA observations will eventually offer frequency dependent data at different frequency bins with high statistical significance that will enable us to quantitatively test modified dispersion relations.

The paper is organized as follows. In Sec.~\ref{sec:review} we briefly review the formalism of the decomposition of ORF into Legendre polynomials in the PTA system.
In Sec.~\ref{sec:coeff}, we give a full analytical expression for the coefficients of these Legendre polynomials. 
It is based on the assumption that distances to the pulsars are much longer than the wavelength. Then we also study the finite distance effect and find that its effect is non-perturbative in the subluminal case.   
In Sec.~\ref{sec:CPTA}, we study the constraints on the phase velocity from CPTA data.   
Throughout the paper we use the metric signature $(-,+,+,+)$, and set $\hbar = c = 1$.

\section{Introduction on the pulsar timing array system} \label{sec:review}
In this section, we review how to obtain a gravitational wave background signal in the PTA system. As mentioned above, each pulsar can be treated as a standard clock, and their pulse frequency variation induced by the underlying metric fluctuations will accumulate in the residual of arrival time, which is the observable for each pulsar. It is defined as 
\begin{equation}
\label{eq,R}
R(t,\hat p) \equiv \int_0^t dt^\prime\, \left(\frac{\nu_0(\hat p)-\nu(t^\prime,\hat p)}{\nu_0(\hat p)}\right) =   \int_0^t dt^\prime z(t^\prime,\hat p)\ ,
\end{equation}
where $\nu$ is the pulse frequency, and the integrand is the difference in the frequencies propagating in the direction of the unit vector $\hat p$ due to the presence of gravitational waves, which we will refer to as the redshift, $z(t, \hat p)$. By measuring the arrival time residual of each individual pulsar, one can construct the two point correlation function of the pulsar pairs. 

We evaluate the redshift by solving the geodesic equation for pulsar light.
In the presence of gravitational waves $h_{ij}$, it is given by 
\begin{equation}
\dfrac{d \nu}{d\lambda}  \simeq \frac{\nu^2}{2} \dfrac{dh_{ij}}{dt}   \hat p^i  \hat p^j \ , \label{eq: geo}
\end{equation}
where $\lambda$ is the affine parameter defined such that $dt/d\lambda$ is equal to the frequency, $\nu$. 
 To solve this, we assume that the gravitational wave signal is approximated as a plane wave solution $h_{ij}(t,\vec{x}) = h_{ij}(t-\hat{\Omega}\cdot\vec{x}/v_p(k))$, where $\hat{\Omega}$ is a unit vector from which gravitational waves propagate and $v_p(k) = \omega(k)/k$ is the phase velocity with respect to the angular frequency of gravitational waves, $\omega(k)$.
Then, in terms of the following relationship of time and space variables,
\begin{equation}
    t = \nu_0 (\lambda-\lambda_e) + t_e   \ , \ \ \hat\Omega \cdot \vec x =\hat\Omega \cdot \vec x_e - \nu_0 (\lambda-\lambda_e)  \hat\Omega\cdot \hat p\,,
\end{equation}
between the pulsar and earth (labeled with $e$), \eqref{eq: geo} is solved as
\begin{equation}
\log\left(\dfrac{\nu}{\nu_0}\right) = \dfrac{v_p~  \hat{p}_i\hat{p}_j }{2\left(v_p+ \hat{\Omega}\cdot\hat{p}\right)}\Delta h_{ij} \rightarrow~  z(t, \hat p) \simeq -\dfrac{v_p ~\hat{p}_i\hat{p}_j }{2\left(v_p + \hat{\Omega}\cdot\hat{p}\right)}\Delta h_{ij} \ ,
\end{equation}
where
\begin{equation}
\Delta h_{ij} \equiv h_{ij}\left(t_p-\dfrac{\hat{\Omega}\cdot\vec{x}_p}{v_p}\right) - h_{ij}\left(t_e-\dfrac{\hat{\Omega}\cdot\vec{x_e}}{v_p}\right) \ ,
\end{equation}
is the difference of the gravitational wave at pulsar and at earth.

 For a gravitational wave background, $z(t,\hat{p})$ needs to be averaged over the direction from which gravitational waves propagate.
It reads
\begin{eqnarray}
 z(t, \hat p)
 &=&   \int \frac{dk}{2\pi} d^2\hat\Omega ~ \frac{  p^i p^j h _{ij}(k,t_0)}{2}      e^{i \omega(k) (t-t_0) } \frac{v_p (k)  }{     \left(v_p(k)+ \odp \right)}  \left(1- e^{-ik L \left(v_p(k)+ \odp \right)  } \right)\ .
\end{eqnarray}
Notice that we apply the Fourier transformation in momentum $k$ instead of frequency, which can be related back through $2\pi f = \omega(k)$. 
The time residual is given by
\begin{eqnarray}\label{eq,residue}
   && R(t, \hat p) = \int_0^t z(t^\prime) dt^\prime \nonumber\\
    &=& \int_{-\infty}^\infty \frac{dk}{2\pi} \int_{S^2} d^2\hat\Omega ~ \frac{  p^i p^j h _{ij}(k,t_0)}{-2i \omega(k)}      \left(e^{i \omega(k) t}-1\right) \frac{v_p (k)  }{     \left(v_p(k)+ \odp \right)}  \left(1- e^{-ik L \left(v_p(k)+ \odp \right)  } \right) \ ,
\end{eqnarray} 
where we have set the emission time of the gravitational wave, $t_0$, to be zero.
We can read off the Fourier transformation, $\tilde R(k,\hat p)$, as
\begin{eqnarray}
    \tilde R(k, \hat p) &=&  \frac{\hat p^i \hat p^j}{-2i \omega(k)} \int_{S^2}d^2\hat\Omega ~ \frac{v_p (k)  }{     \left(v_p(k)+ \odp \right)} \left(1- e^{ - i kL \left(v_p+  \odp\right) }\right) h_{ij} \left(k\right)\nonumber\\
    & = &    \sum_{\ell=2}^\infty \sum_{m= -\ell}^\ell \left(R^E_{(\ell m)}(k,\hat p) a_{(\ell m)}^E(k) + R^B_{(\ell m)}(k,\hat p) a_{(\ell m)}^B(k) \right) \nonumber\\
     &=& \int_{S^2} d^2\hat\Omega \left(R^+ (k,\hat p,\hat \Omega) h ^+(k,\hat \Omega ) +R^\times (k,\hat p,\hat \Omega) h ^\times (k,\hat \Omega) \right) 
\end{eqnarray}
where $R^{E,B}_{(\ell m)}(k,\hat p)$ and $R^{+,\times} (k,\hat p,\hat \Omega)$ are the {\it response functions}. $a_{(lm)}^{E,B}$ and $h ^{+,\times}(k,\hat \Omega ) $ are the amplitudes of gravitational waves in spherical harmonics and polarization basis, respectively.
We defer details of the basis and coefficients transformation \cite{Gair:2014rwa,Liang:2023ary} to Appendix \eqref{app:decomposition}.

The two point correlation functions of the arrival time $R(k,\hat p)$ can be obtained through
\begin{eqnarray}
\label{eq,twopt}
    \braket{R(t, \hat p_1 ) R(t^\prime,\hat p_2)  } = \int_{-\infty}^\infty \frac{dk}{2\pi } e^{  i \omega(k)(t- t^\prime )} H(k) \Gamma(k,\xi)\ ,
\end{eqnarray}
where $H(k)$ is the power spectrum of the GWB defined by 
\begin{align}
&\braket{h^+(k,\hat{\Omega}) h^{+*}(k',\hat{\Omega}')} = \braket{h^\times(k,\hat{\Omega}) h^{\times*}(k',\hat{\Omega}')} = \pi H(k)\delta(k-k')\delta^2(\hat{\Omega}-\hat{\Omega}')\,,
\end{align}
or
\begin{align}
&\left\langle a_{(\ell m)}^E(k) a_{\left(\ell'  m' \right)}^{E *}\left(k' \right)\right\rangle=\left\langle a_{(\ell m)}^B(k) a_{\left(\ell' m' \right)}^{B *}\left(k' \right)\right\rangle = 2\pi\delta_{\ell\ell'}\delta_{mm'}H(k)\delta(k-k') \,,
\end{align}
and $\Gamma(k,\xi)$ is the momentum-dependent overlap reduction function (ORF) of the angle between pairs of pulsars $\xi \equiv \cos^{-1}(\hat{p}_1\cdot\hat{p}_2)$, which contains information about the angular correlation. 

In polarization basis, the overlap reduction function takes the form,
\begin{eqnarray}
\label{eq,gammaint}
    && \Gamma(k,\xi) = \beta\int_{S^2}d^2 \Omega \sum_{A= +,\times } R^A (k,\hat{p}_1,\hat\Omega )  R^{A*}  (k,\hat{p}_2,\hat\Omega )   \ ,
\end{eqnarray}
 where $\beta$ is a normalization factor, and in harmonic basis, it can be expressed as
\begin{eqnarray}
\label{eq,gammasum}
    \Gamma(k,\xi) &=& \mathcal{C} \sum_{\ell=2}^{\infty} \Gamma_{12, \ell}  (k,\xi) = \mathcal{C}  \sum_{\ell=2}^{\infty} \sum_{m= -\ell}^{\ell} \sum_{P= E,B } R_{(\ell m)}^P (k,\hat{p}_1)  R_{(\ell m)}^{P*} (k,\hat{p}_2)  \nonumber\\
    &=& \mathcal{C}\sum_{\ell=2}^{\infty} (2\ell+1)\frac{2 (\ell-2)!}{(\ell+2)!} |c_\ell(k)|^2 P_\ell(\cos\xi)\equiv \mathcal{C}\sum_{\ell=2}^{\infty} a_\ell P_\ell(\cos\xi) \ ,
\end{eqnarray}
where $\mathcal{C}$ is the overall  normalization factor, and the coefficients can be computed as
\begin{eqnarray}
\label{eq,coeff}
    c_\ell (k) 
        &=& \int_{-1}^1 dx \frac{v_p(k)} { v_p(k)+x } \left[1-e^{-i k L\left(v_p +  x\right)}\right] (1-x^2)^2 \frac{d^2}{dx^2 } P_\ell(x )\ .
\end{eqnarray}
In both GR and massive gravity (superluminal phase velocity) cases, the exponential factor in Eq.~\eqref{eq,coeff} can be dropped to a good approximation in the PTA system, where the typical distances between pulsar and earth are kpc, and the frequency band is 1-100 nHz. The coefficients without the exponential term give $c_\ell = 4(-1)^\ell$ in GR, which will give the well-known HD curve when substituted to Eq.~\eqref{eq,gammasum}. 
However, as pointed out in \cite{Liang:2023ary}, the subluminal phase velocity case has to be treated differently. If we neglect the exponential term, the integrand is ill-defined as there appears a singularity at $x=-v_p$, corresponding to a ``surfing'' angle $\theta = -\cos^{-1}(-v_p)$.
Integrating this exponential factor is computationally expensive, but necessary to compare with data to extract useful constraints.  
In the next section, we present an analytical expression for these coefficients.

\section{Computing the Overlap Reduction Function Beyond GR: Examples}\label{sec:coeff}
The main result of this paper is to give an analytical expression for the coefficients $c_\ell$ in Eq.~\eqref{eq,coeff}. After some manipulations, we obtain
\begin{eqnarray}
\label{eq,analytic}
    c_\ell  = \left\{
  \begin{array}{@{}ll@{}}
   -2 v_p (1+\ell) \left(( (2+\ell)v_p^2-\ell)Q_\ell(-v_p)+2v_p Q_{\ell+1} (-v_p)\right) + \mathcal{O}\left(\frac{1}{kL}\right); ~& v_p(k)>1\,, \\
   4(-1)^\ell + \mathcal{O}\left(\frac{1}{kL}\right);~ &v_p(k)=1\,, \\
    -2 v_p (1+\ell) \left(( (2+\ell)v_p^2-\ell)\tilde Q_\ell(-v_p)+2v_p \tilde Q_{\ell+1} (-v_p)\right) \\
    ~+\pi i v_p (1+\ell) \left(( (2+\ell)v_p^2-\ell)P_\ell(-v_p)+2v_p P_{\ell+1} (-v_p)\right)+ \mathcal{O}\left(\frac{1}{kL}\right);~ & v_p(k)<1\,,
  \end{array}\right.\nonumber\\
\end{eqnarray}
where $P_\ell(x)$ is the Legendre P polynomial, and $Q_\ell$ and $\tilde Q_\ell$ are Legendre Q functions defined as
\begin{eqnarray}
\label{eq,Qdefinition}
    Q_\ell (z)= -\frac{1}{2} P_\ell(z) \log \frac{z+1}{z-1}+ \frac{1}{\ell! 2^\ell}\frac{d^\ell}{dz^\ell} \left((z^2-1)^\ell\log \frac{z+1}{z-1} \right)\quad\text{for}\ z\notin [-1,1]\ ,\nonumber\\
   \tilde  Q_\ell (x)= -\frac{1}{2} P_\ell(x) \log \frac{1+x}{1-x}+ \frac{1}{\ell! 2^\ell}\frac{d^\ell}{dx^\ell} \left((x^2-1)^\ell\log \frac{1+x}{1-x} \right)\quad\text{for}\ x\in [-1,1]\ .
\end{eqnarray}
The difference between the two Legendre Q functions is the choice of the branch cut. In our definition, $  Q_\ell(z)$ has branch cut over $[-1,1]$, while $\tilde   Q_\ell(x)$  has branch cut on $(-\infty,-1)\cup(1,\infty)$. 
Note that the choice of the Legendre Q functions in \eqref{eq,analytic} is consistent with the fact that the real part of $c_\ell$ must be real. 
We save the detail of the algebraic derivation to Appendix \eqref{app:legendreQ}. 

\begin{figure}[h!]
    \centering
 \includegraphics[scale=0.5]{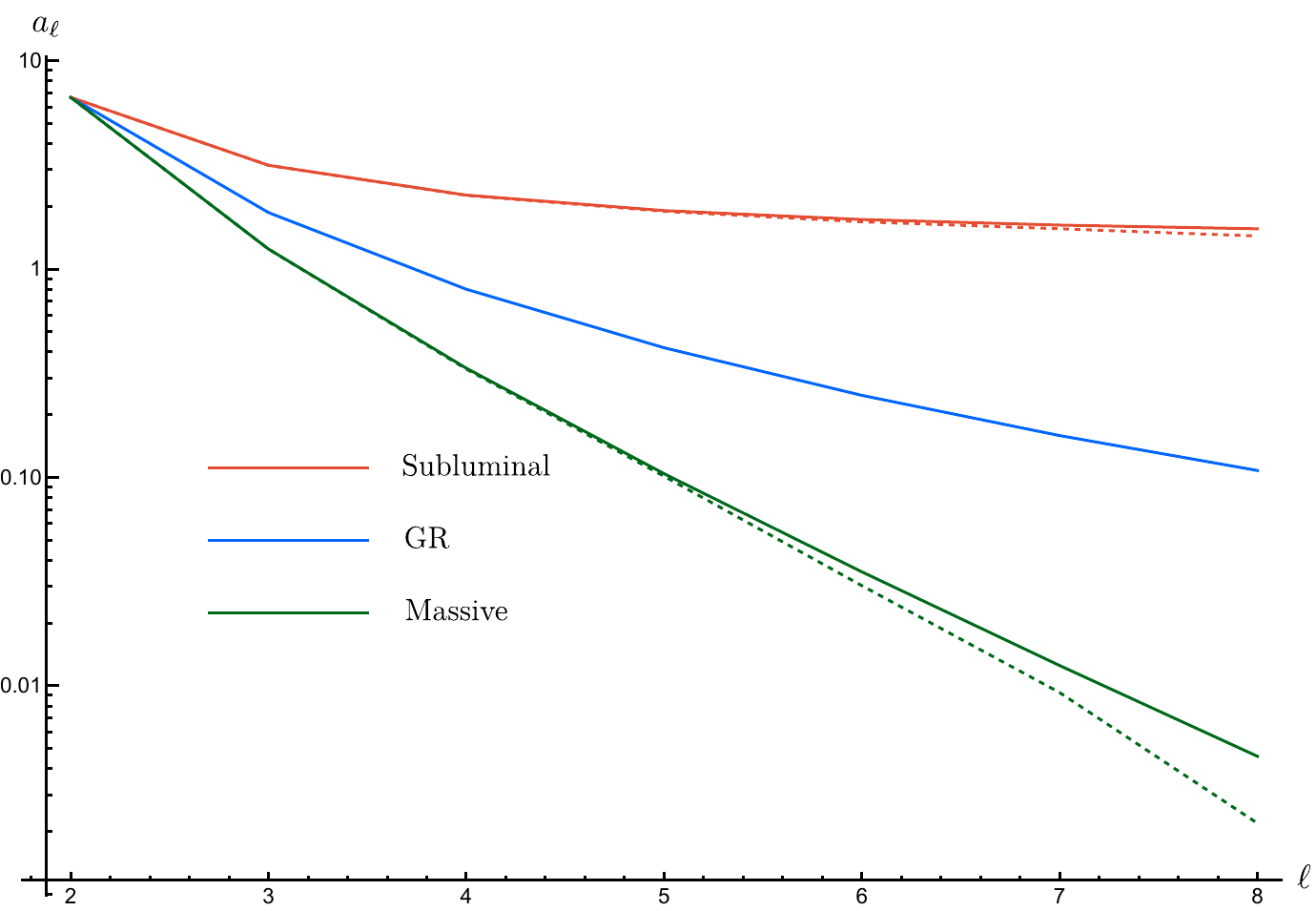}
    \caption{A comparison of the coefficients of the first $8$ Legendre P polynomials $a_\ell$ defined in Eq.~\eqref{eq,gammasum} obtained from numerical integration of Eq.~\eqref{eq,coeff} and from the analytical expression given in Eq.~\eqref{eq,analytic}. 
    The solid lines represent the analytical results where the infinite distance limit $kL\to \infty$ is taken. The red, blue and green colors represent the cases of subluminal phase velocity, general relativity, and massive gravity, respectively. 
    The dashed lines are the numerical results, assuming $kL = 6 $ to exaggerate the finite distance effect. In a real system, $kL$ is usually an order of thousands, therefore, the deviation from our analytical expression to the numerical integration can be neglected.}
    \label{fig:comparison}
\end{figure}

In Fig.~\ref{fig:comparison}, a comparison of the coefficients $a_\ell$ defined in Eq.~\eqref{eq,gammasum} is made between the numerical and analytical results for both massive gravity and subluminal cases.
To show the finite distance effect in the plot, we exaggeratedly chose the parameter $kL = 6$, which is usually the order of $1000$ in a real PTA system. 
One can see that the two match well. 
The numerical result starts to deviate from the analytical expression for $\ell >kL $.   
Thus, we can claim that the analytical expression for the coefficients of the overlap reduction function in harmonic decomposition is a good approximation to the response function for PTA measurements. 

It is worthwhile pointing out that the imaginary part of $c_\ell$ is non-vanishing when $v_p <1$. 
In this case, we can express the real and imaginary parts of the coefficients at the large $\ell$ limit as
\begin{align}
&\lim_{\ell\to\infty} \Re\, c_\ell
 \nonumber\\
 &~= -\pi v_p  \sqrt{\frac{\theta}{\sin \theta}}(\ell+1) \left((-\ell + (2+\ell)v_p^2)Y_0\left(\left(\ell + \frac{1}{2}\right)\theta \right)  +2v_p Y_0\left(\left(\ell + \frac{3}{2}\right)\theta \right)  \right) \,,
\\
  & \lim_{\ell\to\infty}  \Im\, c_\ell 
  \nonumber\\
  &~=   \pi v_p \sqrt{\frac{\theta}{\sin \theta}} (\ell+1) \left((-\ell + (2+\ell)v_p^2)J_0\left(\left(\ell + \frac{1}{2}\right)\theta \right) +2v_p J_0\left(\left(\ell + \frac{3}{2}\right)\theta \right)\right)\, , 
 \end{align}
where $\cos\theta= -v_p$, and $J_0$ and $Y_0$ are the Bessel functions of the 0-th order.
Here, we have used the asymptotic behavior of the Legendre functions for large arguments.
Notice that for $c_\ell = \Re \,c_\ell + i\, \Im\, c_\ell$, we have
\begin{align}
 & \lim_{\ell\to\infty}  c_\ell 
 \nonumber\\
 &~=  i \pi v_p \sqrt{\frac{\theta}{\sin \theta}} (\ell+1) \left((-\ell + (2+\ell)v_p^2)H^{(1)}_0\left(\left(\ell + \frac{1}{2}\right)\theta \right) +2v_p H^{(1)}_0 \left(\left(\ell + \frac{3}{2}\right)\theta \right)\right),
\end{align}
where $H^{(1)}_0$ is the 0-th order Hankel function of the first kind.

Let us focus on the large $\ell$ behavior in the subluminal case.
Using the asymptotic behavior of the Hankel function for large arguments, we obtain 
\begin{eqnarray}
    \lim_{\ell\to\infty} |c_\ell|^2 = 2\pi v_p^2 \frac{1}{\sin\theta} (\ell+1)^2 \Bigg|\frac{(-\ell+ (2+\ell)v_p^2)}{\sqrt{\ell+\frac{1}{2}}} + \frac{2 v_p e^{i\theta}}{\sqrt{\ell+\frac{3}{2}} }\Bigg|^2\,.
\end{eqnarray}
Then, in the limit $kL\to\infty$, the coefficients $a_\ell$ approaches a constant at large $\ell$, 
\begin{eqnarray}
   \lim_{\ell\to\infty} a_\ell =  \lim_{\ell\to\infty}  (2 \ell+1) \frac{2(\ell-2) !}{(\ell+2) !}\left|c_{\ell}(k)\right|^2=  8\pi v_p^2(1-v_p^2)^{3/2}
   +\mathcal{O}(1/\ell)\ ,
\end{eqnarray}
as found in \cite{Liang:2023ary}. 
 One may wonder if this indicates the divergence at $\xi \to 0$. It turns out that this is a fake divergence due to the assumption of infinite distance. When one takes account of the finite distance effect, one can see from Fig.~\ref{fig:comparison} 
 that the numerical calculated coefficients starts to deviate from the analytical prediction at $\ell \sim kL$. 
 It is of interest (and perhaps necessary) to study the finite distance effect in the PTA system. 
 We explicitly calculate the perturbative expansion of Eq.~\eqref{eq,coeff} in $1/(kL)$ in Appendix 
 \eqref{app:legendreQ}, and show that the corrections to the coefficients $a_\ell$ are exactly zero at $\mathcal{O}(1/kL)$ and $\mathcal{O}(1/(kL)^2)$. 
 This indicates that the finite distance effect might be a non-perturbative effect, which gives an exponential decay like $e^{-\ell/kL}$.  
 This exponential decay behavior means a truncation in the summation of the Legendre P polynomials, and eventually leads to a finite result at $\xi \to 0$. 
 
\section{Constraints from current CPTA data}\label{sec:CPTA}
In this section, as an application of our analytical formula, we consider observational constraints on the phase velocity.
We use the data set given by CPTA collaboration since they provide the frequency dependent angular correlation function in their 3.4 years result. Although they provide data sets at three different frequencies, we use the $1.5/3.4 yr$ data since it has the best statistical significance. \footnote{The use of the lowest frequency data is mostly affected by systematic errors due to modelling the pulsar timing. 
On the other hand, the highest frequency also has a low significance due to a suppressed signal in the power law modelling. We expect a future data release may give us freqeuncy-dependent ORFs at various frequencies. }

We use the significance function $\mathcal{S}$ discussed in CPTA collaboration \cite{Xu:2023wog,Jenet:2005pv} to study the statistical significance of the modified Hellings-Downs curve in modified gravity against a constant correlation, 
\begin{eqnarray}
\label{eq,significance}
\mathcal{S} \equiv \sqrt{\frac{N(N-1)}{2}} \frac{\sum_{i<j}\left(c_{i j}-\overline{c}_{i j}\right)\left(\Gamma_{i j}-\overline{\Gamma}_{i j}\right)}{\sqrt{\sum_{i<j}\left(c_{i j}-\overline{c }_{i j}\right)^2 \sum_{i<j}\left(\Gamma_{i j}-\overline{\Gamma}_{i j}\right)^2}}\ ,
\end{eqnarray}
where $i,j$ sum from $1$ to $N$ through all pulsars in the system. $\Gamma_{ij}$ is the theoretical ORF in a modified gravity theory with a pulsar pair $\xi_{ij}=\hat p_i \cdot \hat p_j$ obtained throught Eq.~\eqref{eq,gammasum}, while $c_{ij}$ is the observed angular correlation of this pulsar pair. $\bar c_{ij}$ and $\bar \Gamma_{ij}$ represent the average $\bar c_{ij} = \sum_{i<j} c_{ij}/N_p$, $N_p = N(N-1)/2$ is the number of pulsar pairs in the system.  

\begin{figure}[h!]
    \centering
  \includegraphics[scale=1]{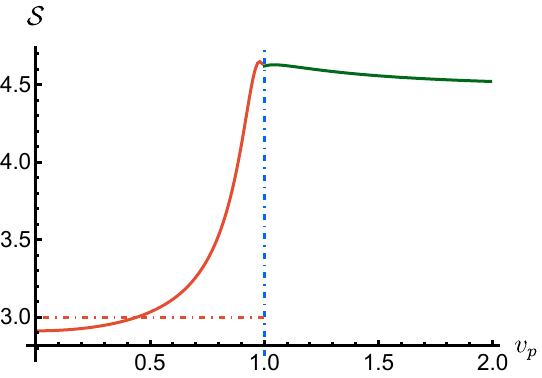}
    \caption{We show the statistical significance of CPTA data as a function of phase velocity. One can see a significant drop down for $v_p <1$. }
    \label{fig:significance}
\end{figure}

We show in Fig.~\ref{fig:significance} the significance function as a function of the phase velocity. 
It is interesting that we can provide quite a good constraint for the subluminal phase velocity case. When $v_p < 0.88$, the significance drops below $4\sigma$, and when the phase velocity approaches $v_p\lesssim 0.4$, the significance drops below $3\sigma$. 
While in the $v_p>1$ case, 
shown in the green curve of Fig.~\ref{fig:significance}, the significance does not decrease rapidly, leaving room for the possibility of massive gravity.

It is also interesting to notice that there are two peaks in the significance function Eq.\eqref{eq,significance}, in both $v_p<1$ and $v_p>1$ regions: $v_p = 0.98$ and $v_p = 1.04$, although they do not have any statistical significance. 
As an exercise, however, let us see if these peaks can be explained by modified gravity.
Let us assume the modified dispersion relation,
\begin{eqnarray}
    \omega^2 = c_s^2 k^2 + m^2\ , v_p = \frac{\omega}{k } = \sqrt{c_s^2 + \frac{m^2}{k^2}}\,.
\end{eqnarray}
Apparently a massless gravity theory with $c_s=0.98$ is a candidate for the peak $v_p=0.98$, though constraints from other observational data would probably exclude such a case. 
If we focus on the peak at $v_p= 1.04$, this corresponds to a graviton mass $m = 1.02\times 10^{-22}$eV if $c_s=1$. It seems this massive gravity theory is still viable \cite{Finn:2001qi,LIGOScientific:2016lio,Liang:2021bct,LIGOScientific:2021sio,Bernardo:2023mxc,Wang:2023div,Wu:2023rib,Bernardo:2023zna,Wu:2023rib,Wu:2023pbt}.
 
\section{Conclusions and Discussions}\label{sec:conclude}
In this paper we gave an analytical expression for the coefficients of the multipole moments of the overlap reduction function for the angular correlation function in the PTA system. 
Then as an application, adopting the recent data from CPTA, which provides a frequency dependent data set, we studied constraints on the phase velocity. 
Although the current data has large uncertainty and could not provide good constraints to the phase velocity, we find it disfavors the subluminal phase velocity case since the significance function drops down drastically.

We conclude that in the future data release, the joint constraints from different frequency bands should be available, and the PTA collaborations should provide the frequency-dependent ORF as part of the results.

\acknowledgments
We thank Kejia Lee for providing CPTA data for us.
We thank Siyuan Chen, Meng-Xiang Lin, Mark Trodden, and Wayne Hu for useful discussions. 
This work is supported by the World Premier International Research Center Initiative (WPI), MEXT, Japan, and also supported in part by JSPS KAKENHI Nos. JP19K14702 (IO), JP20H05859 (IO), JP20H05853 (MS), and JP24K00624 (MS).

 \appendix
 \section{Spherical harmonics decomposition} 
 \label{app:decomposition}
 In this section, we follow \cite{Gair:2014rwa,Liang:2023ary} and list the basis transformation between the polarization basis and the spherical harmonic function basis. 

The harmonic basis $Y^{E,B}_{(lm) ij}(\hat\Omega)$ and the polarization basis $e^{+,\times}_{ij} (\hat\Omega)$ are related through the spin-weighted spherical harmonics $~_{\pm2}Y_{lm} (\hat\Omega)$, 
\begin{eqnarray}
    & Y^{E}_{(lm) ij}(\hat\Omega) = \frac{1}{2} \left(~_2Y_{lm} (e_{ij}^+(\hat\Omega) - ie_{ij}^\times(\hat\Omega)  )+ ~_{-2}Y_{lm} (e_{ij}^+(\hat\Omega) + ie_{ij}^\times(\hat\Omega)  )  \right)\ , \nonumber\\
   & Y^{B}_{(lm) ij}(\hat\Omega) = \frac{1}{2} \left(~_2Y_{lm} (e_{ij}^+(\hat\Omega) + ie_{ij}^\times(\hat\Omega)  )+ ~_{-2}Y_{lm} (e_{ij}^+ (\hat\Omega)- ie_{ij}^\times(\hat\Omega)  )  \right) \ .
\end{eqnarray}
The coefficients of gravitational wave $h_{ij}$ in these two basis are related as
\begin{eqnarray}
    & h^+(k,\hat\Omega) = \frac{1}{2} \sum_{lm}  \left( ~_2Y_{lm} (a_{lm}^E (\hat\Omega)+ ia _{lm}^B(\hat\Omega)  ) +~_{-2}Y_{lm} (a_{lm}^E (\hat\Omega)- ia _{lm}^B(\hat\Omega)  )   \right )\ ,\nonumber\\
 & h^\times (k,\hat\Omega) = \frac{1}{2} \sum_{lm}  \left( ~_2Y_{lm} (a_{lm}^E(\hat\Omega) - ia _{lm}^B(\hat\Omega)  ) +~_{-2}Y_{lm} (a_{lm}^E (\hat\Omega)+ ia _{lm}^B (\hat\Omega) )     \right )\ .
\end{eqnarray}
 
\section{Legendre Q functions}
\label{app:legendreQ}
 We first consider the change of the integration variable from $x$ to $y$,
\begin{equation}
\label{eq,yint}
    y \equiv kL (x+v),\ \ c_\ell = v \int_{kL(v-1)}^{kL(v+1)} dy \frac{\left(1- e^{-i y}\right) }{y} \left[ (1-x^2)^2\frac{d^2}{dx^2} P_\ell (x) \right]_{x= -v+ \frac{y}{kL} } \ ,
\end{equation}
 where we note that 
 \begin{equation}
   (1-x^2)^2\frac{d^2}{dx^2} P_\ell (x)  = (\ell+1) \left((-\ell + (2+\ell)x^2)P_\ell(x) -2x P_{\ell+1} (x)\right)  \, .
   \label{dP2}
 \end{equation}
The exponential factor in Eq.~\eqref{eq,yint} can be split into two parts,
 \begin{equation}
     \frac{1-e^{-i y}}{y} = \frac{1-\cos y}{y} + i\frac{\sin y}{y} \ .
 \end{equation}
 The first term is odd in $y$ while the second term is even. As $kL \to \infty$, the high oscillation behavior for the even term $\sin y/y$ will pick up the phase at $y = 0$, provided that $v<1$. 
 Therefore, the imaginary part of $c_\ell$ is 
 \begin{equation}
     \Im c_\ell = v \pi (\ell+1) \left((-\ell + (2+\ell)v^2)P_\ell(-v) +2v P_{\ell+1} (-v) \right)\ . 
 \end{equation}
 Numerically this matches the result well.
 
 One can also verify that the imaginary part vanishes when $v \geq 1$. 
 This is because, when $v>1$, the phase in the exponent oscillates infinitely many times in the limit $kL\to\infty$, due to the fact that there exists no gravitational wave whose phase can remain constant along the null geodesic from a pulsar to the earth. 
 In the case of $v=1$, i.e., the case of general relativity, the only gravitational wave whose phase can remain constant is the one that propagates along the same null geodesic from the pulsar to the earth, which is an isolated point in the solid angle $\cos^{-1}(\hat{\Omega}\cdot\hat{p})$ and hence does not contribute to the integral. 
 On the other hand, if $v<1$, there exists a circle in the solid angle in the momentum space toward which the gravitational wave phase remains constant. This makes the integral along the circle finite.  
In other words, the imaginary term is due to the coherent sum of the gravitational waves whose propagation directions are toward this circle on which the phase remains constant. 

We now focus on the real part of $c_\ell$. One may naively guess that since $(1-\cos y) / y$ is odd, as $kL\to \infty$, this would vanish under the integration. However, expanding the Legendre polynomial in series of associated Legendre polynomial,
\begin{equation}
    P_\ell \left(-v + \frac{y}{kL}\right) = \sum_{m = 0}^{\ell} \frac{(-1)^m P^m_{\ell} (-v)  }{(1-v^2)^{m/2} m! } \left(\frac{y}{kL}\right)^m \ ,
\end{equation}
we see that it contains both even and odd terms. 
For even $m$, since $(1-\cos y) / y$ is odd, they vanish after integration. 
However, the odd $m$ terms are non-vanishing and they give the real part of the coefficients. 
In principle, one can integrate each term in the series and resum to obtain the result analytically.  
Unfortunately, the expression is too complicated to resum for general $\ell$ since each term contributes at the same order. 

Nevertheless, in the limit, $kL\to\infty$, one can resum the leading order terms to obtain
\begin{eqnarray}
\label{eq,LegendreQ}
    &&\int_{kL(v-1)}^{kL(v+1)}dy  \frac{1-\cos y}{y} P_\ell\left(-v + \frac{y}{kL}\right)\nonumber\\
    &=&\int_{kL(v-1)}^{kL(v+1)}dy \frac{1-\cos y}{y}  P_\ell(-v ) + \int_{kL(v-1)}^{kL(v+1)} \frac{d y}{kL}  \sum_{m=1}^\ell P^m_\ell(-v )(-1)^m \frac{1-\cos y}{m!(1-v^2)^{m/2}} \left(\frac{y}{kL}\right)^{m-1}\nonumber\\
    &=& \log\left(\frac{1+v}{1-v}\right) P_\ell (-v) + \sum_{m=1}^l P^m_\ell(-v )(-1)^m \frac{ (1+v)^m-(v-1)^m  }{m!m (1-v^2)^{m/2}} + \mathcal{O}(1/kL)\nonumber\\
    &=& -2\tilde Q_\ell(-v) + \mathcal{O}(1/kL)\ ,
\end{eqnarray}
where we have used the definitions of the Legendre Q functions in Eq.~\eqref{eq,Qdefinition}. 
We note that, in the limit $kL\to \infty$, there is an interesting similarity of the above to Neumann's integral representation of the Legendre Q functions \cite{bateman_bateman_manuscript_project_1953},
\begin{eqnarray}
    Q_\ell (z)= \frac{1}{2}\int_{-1}^1 \frac{ P_\ell (v) }{z-v}d v\,;~ z\notin [-1,1]\,.
\end{eqnarray}
The difference is that this $Q_\ell(z)$ is defined with a different choice of branch cut (see \eqref{eq,Qdefinition} in the main context).

We can also evaluate the finite distance corrections in Eq.~\eqref{eq,LegendreQ} through the series expansion,  
\begin{eqnarray}
    &&\int_{kL(v-1)}^{kL(v+1)}dy  \frac{1-e^{-i y}}{y} P_\ell\left(-v + \frac{y}{kL}\right) \nonumber\\
    &=&-2\tilde Q_\ell(-v)+ \pi i P_\ell(-v) + \int_{kL(v-1)}^{kL(v+1)} \frac{d y}{kL}   \sum_{m=0}^\ell \frac{ -e^{i y}P^m_\ell(-v )(-1)^m }{m!(1-v^2)^{m/2}} \left(\frac{y}{kL}\right)^{m-1}  \nonumber\\
    &=& -2\tilde Q_\ell(-v)+ \pi i P_\ell(-v)
    \nonumber\\
    &&\quad +  \sum_{m=0}^\ell \frac{  P^m_\ell(-v )(-1)^m }{m!(1-v^2)^{m/2}}  
\Bigl(\gamma\bigl(m,ikL(v+1)\bigr)-\gamma\bigl(m,ikL(v-1)\bigr)\Bigr)\ ,
\end{eqnarray}
where $\gamma(m,z)$ is the incomplete gamma function. It has the asymptotic expansion for large $z$,
\begin{eqnarray}
   \lim_{z\to \infty} \gamma(m,z) = \Gamma(m)- e^{-z}z^{m-1}\left(1+\frac{m-1}{z}+\frac{(m-1)(m-2)}{z^2}+ \mathcal{O}(1/z^3)\right)\ .
\end{eqnarray}
With the above expansion, we could obtain the corrections due to the finite distance effect. We find
\begin{eqnarray}
     &&\int_{kL(v-1)}^{kL(v+1)} dy \frac{\left(1- e^{-i y}\right) }{y} P_\ell\left(-v+ \frac{y}{kL}\right)\nonumber\\
     &&= -2\tilde Q_\ell(-v)+ \pi i P_\ell(-v)+ \frac{i}{kL}\left(\frac{e^{-i kL(v-1)}(-1)^\ell}{v-1}-\frac{e^{-i kL(v+1)}}{v+1} \right)+\mathcal{O}(1/k^2L^2).
\end{eqnarray}
It is easy to see that the $\mathcal{O}(1/kL)$ correction in Eq.\eqref{dP2} vanishes. While the next order $\mathcal{O}(1/(kL)^2)$  is less trivial to see,
one can nevertheless check that their contribution to $c_\ell$ still vanishes. This indicates that the finite distance contribution is perturbatively zero, and therefore, it is a non-perturbative effect.
Deriving this potentially non-perturbative behavior for large $\ell$ is left for future studies.

\bibliography{ref} 
\end{document}